\documentclass[a4paper,english,twocolumn,superscriptaddress,showpacs,aps]{revtex4}
\usepackage[latin1]{inputenc}
\usepackage{graphicx}
\usepackage{amssymb,amsmath}

\begin{document}

\title{Bifurcation and chaos in zero Prandtl number convection}

\author{Pinaki Pal}

\affiliation{Department of Physics, Indian Institute of Technology, Kanpur, India}

\author{Pankaj Wahi}

\affiliation{Department of Mechanical Engineering, Indian Institute of Technology, Kanpur, India}

\author{Mahendra K. Verma}

\affiliation{Department of Physics, Indian Institute of Technology, Kanpur, India}

\author{Supriyo Paul}
\affiliation{Department of Physics, Indian Institute of Technology, Kanpur, India}

\author{Krishna Kumar}
\affiliation{Department of Physics and Meteorology, Indian Institute of Technology, Kharagpur, India}

\author{Pankaj K. Mishra}
\affiliation{Department of Physics, Indian Institute of Technology, Kanpur, India}

\begin{abstract}
We present the detailed bifurcation structure and associated flow patterns near the onset of zero Prandtl number Rayleigh
B\'enard convection.   We employ both direct numerical simulation and a low-dimensional model ensuring qualitative agreement between the two.  Various flow patterns originate from a stationary square  observed at a higher
Rayleigh number through  a series of bifurcations
starting from a pitchfork followed by a Hopf and finally a homoclinic
bifurcation as the Rayleigh number is reduced to the critical value.
Global chaos, intermittency, and crises are observed
near the onset.
\end{abstract}

\maketitle

Thermal convection is observed almost everywhere in the universe: industrial appliances, liquid metals,
atmosphere, oceans, interiors of planets and stars, galaxies etc.  An idealized version of convection
called Rayleigh B\'{e}nard convection (RBC) has been studied for almost a century and it is still an
area of intense research \cite{rbc_etc}.  The two most important parameters characterizing convection in RBC are
the Rayleigh number, describing the vigour of buoyancy, and the Prandtl number, being the ratio of kinetic viscosity and thermal diffusivity.  Solar~\cite{solar} and geological flows~\cite{geo} are considered
to have very low Prandtl numbers, as do flows  of liquid metals~\cite{metal}.  RBC exhibits a wide range of phenomena including instabilities, patterns,  chaos, spatio-temporal chaos, and turbulence for different ranges of Rayleigh number and Prandtl number~\cite{rbc_etc}.  Origin of instabilities, chaos, and turbulence in convection is one of the major research topics of convection.

Direct numerical simulation (DNS), due to its
high dimensionality, generates realistic but excessively voluminous
numerical outputs which obscure the underlying dynamics. Lower
dimensional projections lead to models which, if done improperly,
lose the overall physics. In this letter, our aim is to unfold
and discover the underlying physics of low Prandtl number flows~\cite{lowp} by
examining the natural limit of zero Prandtl number (zero-P)~\cite{thual,spiegel,kumar1,busse,herring,knobloch,pal}. This offers a
dramatic simplification without sacrificing significant physics, as
well as displays a fascinatingly rich dynamic behaviour.
In particular, since zero-P flows are chaotic immediately upon initiation of convection,
we adopt a nonstandard strategy of approaching this system from
the post-bifurcation direction. Moreover, we attack the problem
simultaneously with DNS (to ensure accuracy) as well as a low
dimensional model (to aid physical interpretation); and we
stringently refine both the model and DNS until satisfactory agreement is
obtained at all levels of observed behaviour. Our results show a
diverse variety of both new and previously observed flow patterns.
These flow patterns emerge as a consequence of various bifurcations
ranging from pitchfork, Hopf and homoclinic bifurcations to
bifurcations involving double zero eigenvalues.

Convection in an arbitrary geometry is quite complex, so researchers
have focused on convective flow between two conducting parallel plates
called Rayleigh B\'{e}nard convection~\cite{rbc_etc}.  The fluid has
kinematic viscosity $\nu$,  thermal diffusivity $\kappa$, and
coefficient of volume expansion $\alpha$.  The top and bottom plates are separated by distance $d$ and are
maintained at temperatures $T_1$ and $T_2$ respectively with $T_1 >
T_2$.  Convective flow in RBC is
characterized by  the Rayleigh number
$R = \alpha (T_1 - T_2)g d^3/\nu\kappa$,
where $g$ is the acceleration due to gravity, and the Prandtl number $P = \nu/\kappa$.  Various instabilities, patterns,  and chaos,
are observed for different ranges of $R$ and $P$~\cite{rbc_etc,thual,pattern}. Transition to chaotic states through
various routes have been observed in convection~\cite{expt,dns}.

In this letter, we focus on  zero-P convection.   The governing zero-P Boussinesq equations~\cite{spiegel} are nondimensionalized using $d$ as length scale, $d^2/\nu$ as time scale, and  $\nu (T_1 - T_2)/\kappa$ temperature scale which yields
\begin{eqnarray}
\partial_t(\nabla^2 v_3) &=& \nabla^4 v_3 + R \nabla^2_H \theta \nonumber\\
           & & - \hat{\bf e}_3\cdot\nabla\times
           \left[(\mbox{\boldmath $\omega$}{\cdot}\nabla){\bf v}
           -( {\bf v}{\cdot}\nabla)\mbox{\boldmath $\omega$}  \right],
\label{motion}\\
\partial_t \omega_3 &=& \nabla^2 \omega_3
          +\left[(\mbox{\boldmath $\omega$}{\cdot}\nabla) v_3
          -({\bf v}{\cdot}\nabla)\omega_3\right],
\label{vorticity}\\
  {\nabla}^2 \theta&=& - v_3,\label{energy}\\
\nabla{\cdot}{\bf v} &=& 0, \label{continuity}
\end{eqnarray}
where ${\bf v} \equiv (v_1,v_2,v_3)$ is the velocity field, $\theta$
is the  deviation in the temperature field from the steady conduction
profile, $\mbox{\boldmath $\omega$} = \nabla\times{\bf v}$ is the
vorticity field, $\hat{\bf e}_3 $ is  vertically directed unit
vector, and $ \nabla_H^2 = \partial_{xx} + \partial_{yy}$ is the
horizontal Laplacian.  We consider {\em perfectly conducting
and free-slip boundary} conditions at the top and bottom plates, and
periodic boundary conditions along horizontal directions~\cite{thual,dns}.
In the following discussion we also use the reduced Rayleigh number $r=R/R_c$ where $R_c$ is the
critical Rayleigh number.

Straight two-dimensional (2D) rolls that have zero vorticity are neutrally stable solution of zero-P convection at $r=1$.
However they become unstable for $r>1$.   Busse~\cite{busse}, Thual~\cite{thual} and Kumar et.
al.~\cite{kumar1} showed that these 2D rolls saturate through
generation of vorticity (wavy rolls) for $r>1$ both for low Prandtl number and zero-P fluids.  Thus vorticity plays a critical role in zero-P convection.

Herring~\cite{herring} was first to simulate these equations under
free-slip boundary conditions. However he observed  divergence of
the solutions possibly due to the instabilities described above. The
first successful simulation of zero-P equations with free-slip
boundary conditions was done by Thual~\cite{thual}. He reported many
interesting flow patterns including relaxation oscillation of square
patterns (SQOR) and stationary square patterns (SQ).  Later
Knobloch~\cite{knobloch} explained the stability of the SQ patterns
using the amplitude equations.  Pal and Kumar~\cite{pal} explained the
mechanism of selection of square patterns using a 15-dimensional
model.  Note that asymmetric squares (referred to as `cross roll' in literature) and other patterns have been observed in experiments of low-Prantl number convection~\cite{expt_pattern, rbc_etc}.

We performed around 100 DNS runs of zero-P convection (Eqs.~(\ref{motion}-\ref{continuity}))
using a pseudo-spectral code for
various $r$ values on $64^3$ box. The aspect ratio of our
simulation is $2\sqrt{2} : 2\sqrt{2} : 1$.
In DNS we
observe stationary squares (SQ),  stationary asymmetric squares (ASQ), oscillatory asymmetric squares (OASQ),
relaxation oscillations with squares (SQOR), and chaos.

For our bifurcation analysis we construct a low-dimensional model using the energetic modes of
the above-mentioned simulation in the range of
$r = 1 - 1.4$. We pick 9 large-scale vertical velocity modes (real
Fourier amplitudes): $W_{101}$, $W_{011}$, $W_{202}$, $W_{022}$,
$W_{103}$, $W_{013}$,  $W_{112}$, $W_{121}$, $W_{211}$, and 4
large-scale vertical vorticity modes (real Fourier amplitudes):
$Z_{110}$, $Z_{112}$, $Z_{121}$, $Z_{211}$. The three subscripts are the indices of wavenumber along x,y, and z directions.
Cumulative energy contained in these modes ranges from 85\% to 98\%
 of the total energy of DNS, and each of these modes has 1\% or more of the total energy.
 We derive the model equations
by the Galerkin projection of  Eqs.~(\ref{motion}-\ref{continuity}) on the subspace of these
modes. This results in thirteen coupled first-order ordinary differential equations for the  above variables.  The low-dimensional model captures all the flow patterns of DNS mentioned above. The range of $r$ for these
patterns for the model and DNS are shown in Table~\ref{tab:range_r_patterns},
and they are reasonably close to each other.
Interestingly, the stable steady values of the modes $W_{101}$,
$W_{011}$, $W_{112}$, $W_{121}$, $W_{211}$ for SQ and ASQ patterns
match with corresponding DNS values within 10\%.
\begin{table}[h]
\begin{center}
\begin{tabular}{|c|c|c|}
\hline
Flow patterns & r (Model) & r (DNS)\\
\hline\hline
Chaotic & 1 - 1.0045 & 1 - 1.0048 \\
\hline
SQOR & 1.0045 - 1.0175 & 1.0048 - 1.0708 \\
\hline
OASQ & 1.0175 - 1.0703 & 1.0709 - 1.1315 \\
\hline
ASQ & 1.0703 - 1.2201 & 1.1316 - 1.2005 \\
\hline
SQ & 1.2201 - 1.4373 & 1.2006 - 1.4297 \\
\hline
\end{tabular}
\end{center}
\caption {Range of reduced Rayleigh number $r$ corresponding to various flow patterns observed in the model and the DNS.
Here SQ, ASQ, OASQ, and SQOR represent stationary squares, stationary asymmetric squares, oscillatory asymmetric squares, and relaxation oscillation of squares  respectively. }
\label{tab:range_r_patterns}
\end{table}

The origin of the above flow patterns can be nicely understood using the bifurcation diagram of the low-dimensional model.  To generate the bifurcation diagram, we first evaluate a fixed point numerically using Newton-Raphson method for a given $r$.  The branch of fixed points is subsequently obtained using a fixed arc-length
based continuation scheme~\cite{wahi}. Stability of the fixed
points is ascertained through an eigenvalue analysis of the Jacobian
and accordingly the bifurcation points are located. New branches of fixed points are born when the eigenvalue(s) become zero (pitchfork),
 and branches of periodic solutions appear when the eigenvalue(s) become  purely imaginary (Hopf).   Subsequent
branches are generated by calculating and continuing the new steady
solutions close to the bifurcation points.

\begin{figure}[h!]
\includegraphics[height=7cm,width=8.5cm]{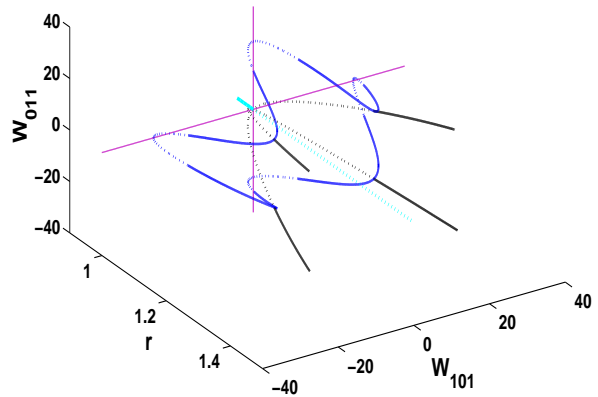}
\caption{Three dimensional view of the bifurcation diagram showing the fixed points with solid and dashed curves representing the stable and unstable fixed points respectively. Black, blue, and cyan curves represent stationary squares (SQ), asymmetric stationary squares (ASQ), and conduction state respectively. All the points on the axis (purple lines) are 2D roll solutions.}\label{fig:3d_bifurcation}
\end{figure}

Fixed points are the backbone of bifurcation diagram.
For $r<1$, the origin is the unique stable fixed point corresponding to
the pure conduction state. There is a double zero eigenvalue at $r=1$~\cite{Guck_Holmes}, and all the fixed points (13 in number) arising from $r=1$ are unstable for $r>1$. These fixed points are
shown as dotted lines in Fig.~\ref{fig:3d_bifurcation}. Four of these branches of
fixed points bifurcate from the origin; these fixed points satisfy
$|W_{101}| =  |W_{011}|$.  The other 8  branches of unstable fixed
points emerge from nonzero $W_{101}$ or $W_{011}$, and they obey
$|W_{101}| \ne |W_{011}|$ (see Fig.~\ref{fig:3d_bifurcation}).  With
an increase of $r$, these 8 branches become stable and merge with
the 4 branches that originate from the origin.  In Fig.~\ref{fig:3d_bifurcation}, modes
$W_{101}$ and $W_{011}$ are presented even though all other modes are also nonzero.

\begin{figure}[h!]
\includegraphics[height=!,width=8.5cm]{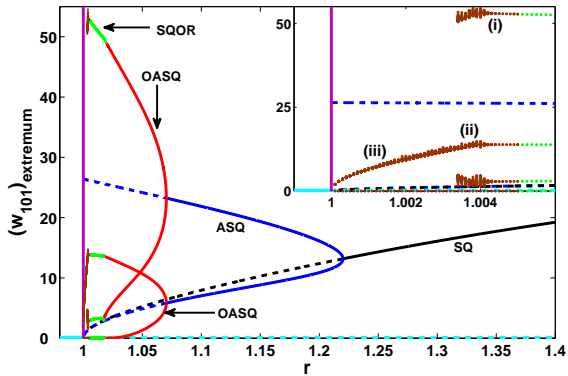}
\caption{Bifurcation diagram of the model for $ 0.95 \le r \le 1.4 $. The stable branches corresponding to stationary squares (SQ) and  stationary asymmetric squares (ASQ)  are represented by  solid black and solid blue lines respectively.  Red, green, and brown curves represent the extrema of oscillatory asymmetric squares (OASQ), relaxation oscillation of squares (SQOR), and chaotic solutions respectively.  A zoomed view of the bifurcation diagram for the chaotic regime is shown in the inset.  Branches corresponding to the unstable fixed points are represented by dashed lines. Cyan line represents the conduction state.}\label{fig:bifurcation}
\end{figure}

After a discussion on fixed points, we focus on the complete bifurcation diagram shown in Fig.~\ref{fig:bifurcation}.
 Chaotic solutions are observed at the
onset of convection itself, i.e., just above $r=1$.  A better insight into the origin of the various solutions is facilitated by starting the analysis at a higher $r$ value and tracking the various bifurcations while approaching $r=1$.

\begin{figure}[h!]
\includegraphics[height=!,width=8.5cm]{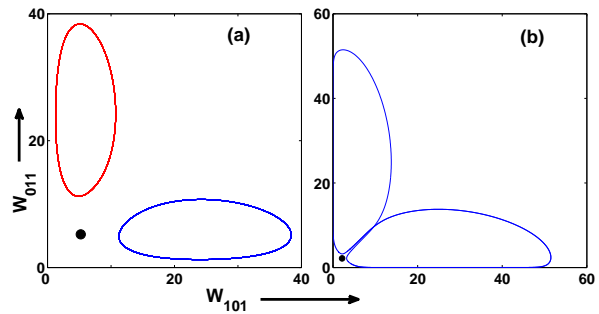}
\caption{Phase space projections of limit cycles on the $ W_{101} - W_{011} $ plane for (a) $r = 1.0494$ and (b) $r = 1.0099$. The limit cycles in figure (a) merge to form a single limit cycle in figure (b).   Black dots indicate the symmetric square saddle.}\label{fig:limit_cycle}
\end{figure}

We start our analysis at $r=1.4$ where we observe stable symmetric
squares ({SQ}) with $|W_{101}| =  |W_{011}|$  (black curve in
Fig.~\ref{fig:3d_bifurcation}).  In Fig.~\ref{fig:bifurcation} we
represent only the $W_{101} =  W_{011}$ solution.   As $r$ is
reduced from $1.4$, the SQ branch of fixed points loses stability
via a supercritical pitchfork  bifurcation at $r \approx 1.2201$,
after which we observe stationary solutions with
$W_{101} \ne  W_{011}$ (blue curves of
Figs.~\ref{fig:3d_bifurcation} and \ref{fig:bifurcation}). These
solutions correspond to asymmetric square patterns (ASQ), either
dominant along x axis  ($|W_{101}| >  |W_{011}|$), or dominant along
y axis ($|W_{101}| <  |W_{011}|$).  The SQ solution $|W_{101}| =
|W_{011}|$ continues as a saddle.   With a further reduction of $r$,
ASQ branches lose stability through a supercritical Hopf  bifurcation
at $r \approx 1.0703$ and limit cycles are born. These limit cycles are represented by red curves in
Fig.~\ref{fig:bifurcation}.  Physically they represent oscillatory
asymmetric square patterns (OASQ).   Fig.~\ref{fig:limit_cycle}(a)
illustrates the projection of two of these stable limit cycles (for
$r=1.0494$) on the $W_{101}-W_{011}$ plane.

The limit cycles grow in size as $r$ is lowered.  A homoclinic orbit is formed at  $r \approx 1.0175$.
Afterwards,  homoclinic chaos is observed in a narrow window.  At lower $r$ the
attractor becomes regular resulting in a larger limit cycle that
corresponds to the relaxation oscillations with an intermediate
square pattern (SQOR).  Fig.~\ref {fig:limit_cycle}($b$) illustrates
the projection of this limit cycle at $r=1.0099$.  The flow
pattern in this regime changes from an approximate pure roll in one
direction to a symmetric square, and then to an approximate pure roll
in the perpendicular direction.   The SQOR solution
is represented by the green curve  in Fig.~\ref{fig:bifurcation}.

\begin{figure}[h!]
\includegraphics[height=!,width=7.2cm]{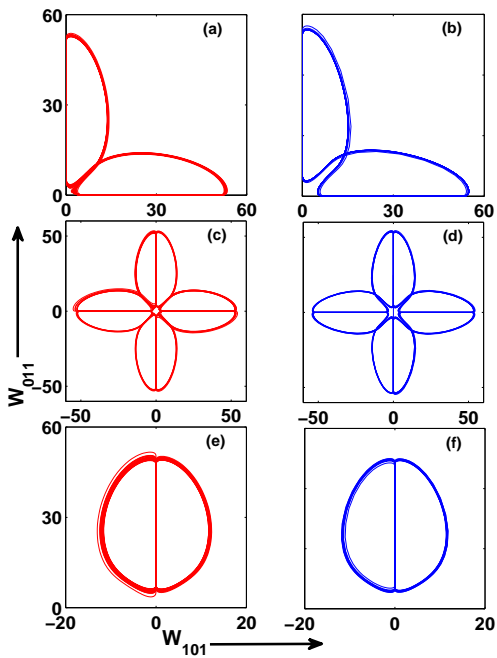}
\caption{The three different chaotic solutions observed  near $r=1$:  Ch1 at $r = 1.0041$ for the model (a) and  at $r=1.0045$ in DNS (b);
Ch2 at $r = 1.0038$ for the model (c) and at $r=1.0030$ in DNS (d);   Ch3 at $r = 1.0030$ for the model (e) and at $r = 1.0023$ in DNS (f).  These solutions belong to (i), (ii), and (iii) regimes in the bifurcation diagram (Fig.~\ref{fig:bifurcation}).}\label{fig:chaos}
\end{figure}

The flow becomes chaotic as $r \rightarrow 1$. The chaotic flow manifests itself
in three different forms: Ch1, Ch2, and Ch3 as shown in the inset of
Fig.~\ref{fig:bifurcation} as (i), (ii), and (iii) respectively. The
phase space projection for these three solutions are depicted in
Fig.~\ref{fig:chaos} for $r=1.0041$, 1.0038 and 1.0030 for the 13-mode model, and for $r=1.0045$, 1.0032 and 1.0023 in the DNS.
Their chaotic nature is confirmed by the positivity of
the largest Lyapunov exponent (0.0131, 0.0254 and 0.0389) calculated using the 13-mode model for $r=1.0041$, 1.0038 and 1.0030 respectively.

The first chaotic solution, Ch1
(Figs.~\ref{fig:chaos}(a) and~\ref{fig:chaos}(b)) results from the broadening of the limit
cycle attractor with chaotic switchings between the two lobes of the
attractor. This global chaos could probably be attributed to
homoclinic tangles.  The time-series of the solution shows
intermittency. At  $r \approx 1.004$, these four chaotic attractors of Ch1 merge in a
`crisis' to yield a single large chaotic attractor Ch2
(Figs.~\ref{fig:chaos}(c) and~\ref{fig:chaos}(d)).
This chaotic solution  persists till $r \approx 1.0035$ after which
it splits into four different chaotic attractors Ch3 in another
`crisis', one of which is shown in Figs.~\ref{fig:chaos}(e) and~\ref{fig:chaos}(f).  The time-series again
shows intermittency, and the flow pattern switches from an
approximately pure roll in one direction to an asymmetric square.
With a further reduction
in the Rayleigh number, the size of
these chaotic attractors decreases and they ultimately merge with
one of the branches of the unstable ASQ fixed points at $r=1$. In
Fig.~\ref{fig:bifurcation}, we exhibit the merger of one of these
chaotic attractors with the unstable ASQ fixed point with $W_{101}
\rightarrow 0$.

In conclusion, we present for the first time a numerically obtained, DNS validated, detailed bifurcation diagram and associated flow structures of zero-P convective flow near the onset of convection. The whole spectrum of phenomena
observed in DNS near the onset of convection is replicated by the low-dimensional
model.  Hence, the bifurcation structure presented here explains the origin and dynamics
of various patterns near the onset of convection.
Recent analysis of VKS (Von-Karman-Sodium) experimental results
indicate a strong role of large-scale modes for the magnetic field reversal~\cite{VKS}.
A study of large-scale modes as outlined in this letter may provide
useful insights into the mechanism behind the  generation and reversal of magnetic field.
The dynamics of large-scale modes in other hydrodynamic systems like rotating turbulence,
magneto-convection etc.~could also be captured by a similar approach.

A careful analysis of DNS results indicate the existence of fringe attractors apart from the main attractors presented
in this letter.  These new attractors are relatively
insignificant and they exist only in localized regimes of $r$. A
detailed investigation of all the attractors will be reported elsewhere.
Bifurcation analysis for $r>1.4$ is reasonably complex as well.
Also preliminary results show a reasonable amount of
similarity between low Prandtl number convection and zero-P
convection.  These issues are under investigation.

We thank S. Fauve, A. Chatterjee, A. K. Mallik, and V. Subrahmanyam for useful
discussions.  We thank Computational Research Laboratory, India for providing us access to the supercomputer EKA where part of this work was done.
This work was supported by  the research grant of Department of Science and Technology,
India.

%
%
%

\end{document}